\documentclass[%
reprint,
 amsmath,amssymb,
 aps,
 twocolumn,
prl,
]{revtex4}

\usepackage{graphicx}
\usepackage{dcolumn}
\usepackage{bm}
\usepackage{braket}
\usepackage[
colorlinks=true,
linkcolor=blue,
filecolor=blue,
citecolor=blue,
urlcolor=blue,
]{hyperref}

\newcommand{\blue}[1]{\textcolor{blue}{#1}}

\usepackage{comment}
\usepackage{siunitx}
\usepackage{color}

\begin{document}
\preprint{APS/123-QED}

\title{{Time-Domain Universal Linear-Optical Operations \\ for Universal Quantum Information Processing}}

\author{Kazuma Yonezu}
\affiliation{%
    Department of Applied Physics, School of Engineering, The University of Tokyo,
    7-3-1 Hongo, Bunkyo-ku, Tokyo 113-8656, Japan
}%
\author{Yutaro Enomoto}%
\affiliation{%
    Department of Applied Physics, School of Engineering, The University of Tokyo,
    7-3-1 Hongo, Bunkyo-ku, Tokyo 113-8656, Japan
}%
\author{Takato Yoshida}%
\affiliation{%
    Department of Applied Physics, School of Engineering, The University of Tokyo,
    7-3-1 Hongo, Bunkyo-ku, Tokyo 113-8656, Japan
}%
\author{Shuntaro Takeda}%
\email{takeda@ap.t.u-tokyo.ac.jp}
\affiliation{%
    Department of Applied Physics, School of Engineering, The University of Tokyo,
    7-3-1 Hongo, Bunkyo-ku, Tokyo 113-8656, Japan
}%

\date{\today}

\begin{abstract}
    {
        We demonstrate universal and programmable three-mode linear optical operations in the time domain by realizing a scalable dual-loop optical circuit suitable for universal quantum information processing (QIP).
        The programmability, validity, and deterministic operation of our circuit are demonstrated by performing nine different three-mode operations on squeezed-state pulses, fully characterizing the outputs with variable measurements, and confirming their entanglement. Our circuit can be scaled up just by making the outer loop longer and also extended to universal quantum computers by incorporating feedforward systems. Thus, our work paves the way to large-scale universal optical QIP.
    }
\end{abstract}

\maketitle
{
    Optics has been crucial in implementing various quantum information processing (QIP), such as quantum computing~\cite{knill2001scheme,Takeda2019Toward}, quantum networking~\cite{wehner2018quantum}, and quantum simulation~\cite{flamini2018photonic}.
    A core technology for universal optical QIP in both qubits and continuous variables is linear optical operations, which linearly transform creation operators of photons~\cite{kok2007linear}.
    Such operations are implementable only with linear optics and can create entanglement between optical modes~\cite{vanLoock2007Building},
    thereby providing core processing functions.
    These functions realize universal QIP, namely an arbitrary unitary operation for either continuous-variable or qubit scheme,
    when combined with appropriate quantum light sources, detectors, and feedforward systems~\cite{knill2001scheme,Takeda2019Toward}.
    Even without the feedforward, linear optical operations allow for implementing
    non-universal QIP, such as boson sampling~\cite{aaronson2011computational, he2017time} and quantum walk~\cite{aharonov1993quantum, schreiber20122d}.
}

{
    Much effort has been devoted to scaling up universal linear optical operations toward large-scale universal quantum processors.
    Thus far, universal linear optical operations have been implemented up to 20 modes by developing multi-mode linear interferometers on programmable photonic chips~\cite{carolan2015universal, Qiang2018Large, taballione202220, taballione2021universal}. In such implementations, one optical path represents one mode, and spatial arrays of phase shifters (PSs) and beam splitters (BSs) perform the desired operations. In this path encoding, increasing the number of modes requires quadratically growing numbers of BSs and PSs. This makes the interferometer larger, and makes the stabilization, calibration, and control of all the interferometric points more difficult, possibly limiting scalability.
}

{
    A more scalable option to realize large-scale linear optical operations is to use temporal encoding, where a large number of modes can be defined as sequential optical pulses on a single optical path~\cite{Takeda2019Toward}.
    High scalability of the temporal encoding has already been shown in recent optical demonstrations of quantum supremacy~\cite{madsen2022quantum}, scalable entanglement generation~\cite{Yokoyama2013Ultra, takeda2019demand, meyer2022scalable, istrati2020sequential}, and multi-mode multi-step quantum gates~\cite{Asavanant2021Time, Larsen2021Deterministic, enomoto2021programmable}. The temporal encoding is also advantageous for scaling up universal linear optical operations by adopting a dual-loop optical circuit proposed in Ref.~\cite{motes2014scalable}.
    Moreover, such a dual-loop architecture is extendable to universal QIP by appropriately incorporating feedforward systems~\cite{Rohde2015Simple,takeda2017universal}.
    Thus far, such architectures have been partly adopted to scale up specific non-universal QIP tasks, such as boson sampling~\cite{he2017time} and quantum walk~\cite{schreiber20122d}.
    However, these experiments were designed for specific sampling tasks and insufficient for universal QIP. More specifically, the linear optical operations in these experiments were not universal due to the lack of complete dynamic controllability of the loops. Moreover, the loops were not phase-stabilized, eliminating the coherence between optical pulses inside and outside the loops. In addition, these experiments only post-select the output to evaluate the sampling tasks and did not confirm the deterministic operation of even the most basic function (e.g. entanglement generation) of the linear optical operations for universal QIP.
}

\begin{figure*}[htbp]
    \centering
    \includegraphics{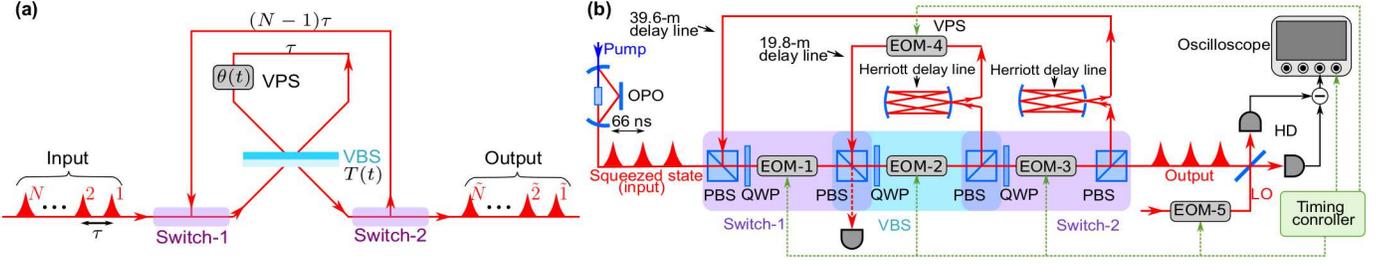}
    \caption{{{Dual-loop circuit for universal linear optical operations in the temporal encoding.}}
            {(a)} {Conceptual schematic.}
            {(b)} {Experimental setup.}
        See text for details. OPO, optical parametric oscillator; PBS, polarizing beam splitter; QWP, quarter-wave plate; EOM, electro-optic modulator.}
    \label{fig:loop}
\end{figure*}

\begin{figure*}[htbp]
    \centering
    \includegraphics[scale = 0.9]{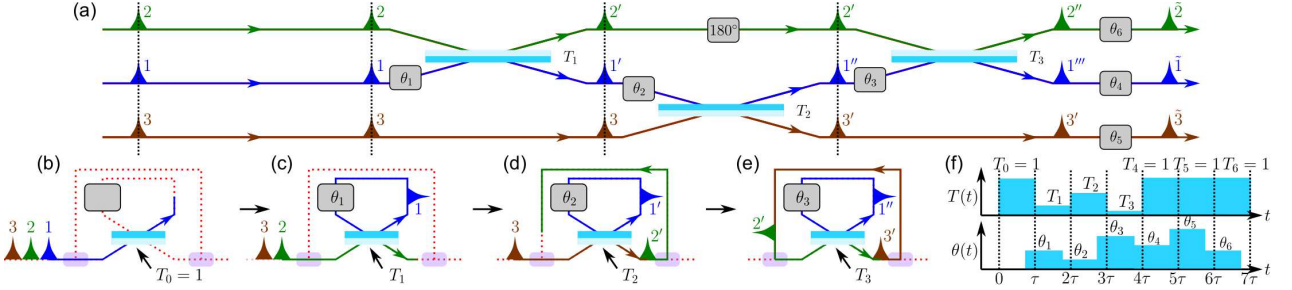}
    \caption{{Dynamics of the dual-loop circuit for universal three-mode linear optical operations.}
            {(a)} One of the possible configurations to perform universal three-mode linear optical operations in the path encoding.
        The sides of the BSs that invert the phase of the reflected modes are colored light blue.
        A phase shift of \ang{180} is added to make the circuit completely equivalent to our dual-loop circuit.~\cite{supp}
        {(b)--(e)} Dynamics of the dual-loop circuit to perform universal three-mode operations in the temporal encoding. Subsequently, the transmissivity of the VBS is kept at 1 and the VPS adds phase shift $\theta_4$, $\theta_5$, and $\theta_6$ to each mode.
            {(f)} Temporal control sequence for the dual-loop circuit.
    }
    \label{fig:3mode_ver2}
\end{figure*}

{
Here, we demonstrate universal and programmable three-mode linear optical operations in the time domain by realizing a scalable dual-loop optical circuit suitable for universal QIP.
Our dual-loop circuit achieves universal linear optical operations by completing all the functionalities in the original proposal~\cite{motes2014scalable}, including a variable beam splitter (VBS), a variable phase shifter (VPS), and fully phase-stabilized dual loops.
We evaluate the performance of our circuit by using a squeezed light source and a homodyne detector with a programmable measurement basis.
The programmability, validity, and deterministic operation of our circuit are demonstrated by performing nine different three-mode operations on squeezed-state pulses, fully characterizing their output states via homodyne detection, and confirming their entanglement.
These results together show the applicability of our circuit to arbitrary input states in both qubit and continuous-variable regimes, leading to universal QIP.
In fact, the extension of our circuit to a universal quantum processor is straightforwardly possible by incorporating the feedforward system already realized in the previous work~\cite{enomoto2021programmable}.
Note that our circuit is designed for various QIP and can process externally injected input states and export the output states, while the previously demonstrated one-way quantum computing circuit~\cite{Larsen2021Deterministic} was designed for computational purposes and internally prepared input states and returned only calculation results instead of output quantum states.
Furthermore, our dual-loop circuit can be straightforwardly scaled up just by making the outer loop longer and storing more modes in the loop.
Thus, our work paves the way to large-scale universal QIP in the time domain.}

\textit{Working principle of the dual-loop circuit} ---In the typical path encoding, universal $N$-mode linear optical operations can be performed by spatial arrays of BSs and PSs~\cite{reck1994experimental}. In the temporal encoding, the same operations can be done by the dual-loop circuit in {Fig.~\ref{fig:loop}\blue{(a)}}~\cite{motes2014scalable}. The working principle of the dual-loop circuit is the following. First, $N$ sequential pulsed optical modes with time interval $\tau$ are injected and stored in the dual-loop circuit via optical switches (Switch-1, 2). Here, $N-1$ modes are stored in the outer loop whose round-trip time is $(N-1)\tau$, while the remaining mode is stored in the inner loop whose round-trip time is $\tau$. The inner loop includes a VBS with transmissivity $T(t)$ and a VPS with phase $\theta(t)$, where $t$ denotes time. This inner loop repeatedly performs two-mode BS interactions between the pulsed modes in the inner and outer loops while dynamically changing $T(t)$ and $\theta(t)$ for each pulse. It can be shown that such operations enable an arbitrary linear optical operation between the $N$ modes~\cite{supp}. After the desired operations, Switch-2 sequentially exports the output modes. This dual-loop circuit is highly scalable since it can process an arbitrary number of modes with a constant number of optical components just by making the outer loop appropriately long. Furthermore, operations are fully programmable since they are determined by the electric control sequence of $T(t)$ and $\theta(t)$.

Figure~\ref{fig:3mode_ver2} exemplifies a more concrete sequence to perform an arbitrary linear optical operation for $N=3$ modes, which we adopt in our experiment. Figure~\ref{fig:3mode_ver2}\blue{(a)} illustrates one of the possible configurations to perform an arbitrary three-mode linear optical operation in the path encoding. The same operation can be done in the dual-loop circuit as shown in Figs.~\ref{fig:3mode_ver2}\blue{(b)-(e)} based on the control sequence in Fig.~\ref{fig:3mode_ver2}\blue{(f)}.
A more general procedure to perform $N$-mode linear optical operations are shown in {Supplemental Material}~\cite{supp}.

\begin{figure*}[htbp]
    \centering
    \includegraphics[scale=0.95]{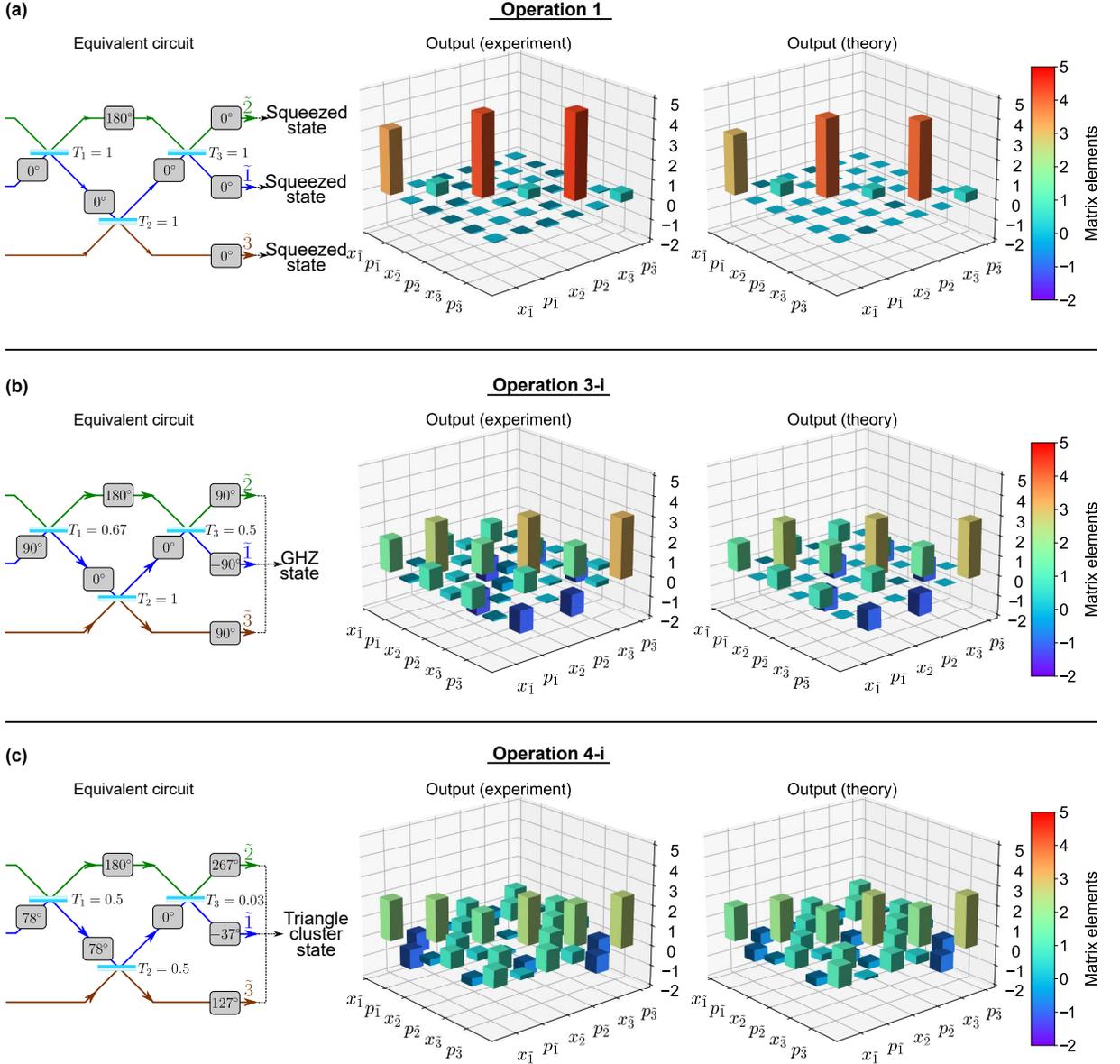}
    \caption{{Representative results of three-mode linear operations in the dual-loop circuit.}
    The matrix elements represent covariances
    $\langle \hat{\xi}_i \hat{\xi}_j + \hat{\xi}_j \hat{\xi}_i \rangle/2 - \langle \hat{\xi}_i \rangle \langle \hat{\xi}_j\rangle$,
    where $\braket{\cdots}$ denotes the mean value and
    $\hat{\xi}=(\hat{x}_{\tilde{1}}, \hat{p}_{\tilde{1}}, \hat{x}_{\tilde{2}}, \hat{p}_{\tilde{2}}, \hat{x}_{\tilde{3}}, \hat{p}_{\tilde{3}})^T$.
    The vacuum variance is set to 1 ($\hbar = 2$).
    Note that the phase-inverting side of one of the three BSs is flipped in (c) since the phase-inverting side of the VBS is flipped when $T<0.5$~\cite{takeda2019demand}.
    See text for details of each column.}
    \label{fig:result_all}
\end{figure*}

\begin{table*}[htbp]
    \centering
    \hspace{-0.5cm}
    \begin{tabular}{c|c|c|c|c}
        Operation                                                     & Output state      & Fidelity($\mathrm{I}$) & Fidelity($\mathrm{I}\hspace{-1.2pt}\mathrm{I}$)                                                                                                                                            & Inseparability parameter \\ \hline \hline
        $1$                                                           &
        \begin{tabular}{c}
            Individual squeezed vacuum \\
            states $(\tilde{1}$, $\tilde{2}$, and $\tilde{3})$
        \end{tabular}                                     & $0.992 \pm 0.002$ & $0.949 \pm 0.003$      & --                                                                                                                                                                                                                    \\\hline
        $2$-i                                                         &
        \begin{tabular}{c}
            EPR state $(\tilde{1}$ and $\tilde{3})$, \\
            Squeezed vacuum state $(\tilde{2})$
        \end{tabular}
                                                                      & $0.958 \pm 0.007$ & $0.894 \pm 0.006$      & $\langle \lbrack \Delta (\hat{x}_{\tilde{1}} - \hat{x}_{\tilde{3}}) \rbrack ^2\rangle+ \langle \lbrack \Delta (\hat{p}_{\tilde{1}} + \hat{p}_{\tilde{3}}) \rbrack ^2\rangle=2.38 \pm 0.05$                            \\\hline
        $2$-ii                                                        &
        \begin{tabular}{c}
            EPR state $(\tilde{2}$ and $\tilde{3})$, \\
            squeezed vacuum state $(\tilde{1})$
        \end{tabular}
                                                                      & $0.966 \pm 0.008$ & $0.907 \pm 0.008$      & $\langle \lbrack \Delta (\hat{x}_{\tilde{2}} - \hat{x}_{\tilde{3}}) \rbrack ^2\rangle+ \langle \lbrack \Delta (\hat{p}_{\tilde{2}} + \hat{p}_{\tilde{3}}) \rbrack ^2\rangle=2.09 \pm 0.03$                            \\\hline
        $2$-iii                                                       &
        \begin{tabular}{c}
            EPR state $(\tilde{1}$ and $\tilde{2})$, \\
            squeezed vacuum state $(\tilde{3})$
        \end{tabular}

                                                                      & $0.965 \pm 0.004$ & $0.896 \pm 0.005$      & $\langle \lbrack \Delta (\hat{x}_{\tilde{1}} - \hat{x}_{\tilde{2}}) \rbrack ^2\rangle+ \langle \lbrack \Delta (\hat{p}_{\tilde{1}} + \hat{p}_{\tilde{2}}) \rbrack ^2\rangle=2.56 \pm 0.03$                            \\\hline
        $3$-i                                                         &
        GHZ state  $(\tilde{1}$, $\tilde{2}$, and $\tilde{3})$        & $0.947 \pm 0.012$ & $0.896 \pm 0.009$      &
        \begin{tabular}{c}
            $\langle \lbrack \Delta (\hat{x}_{\tilde{1}} - \hat{x}_{\tilde{2}}) \rbrack ^2\rangle+ \langle \lbrack \Delta (\hat{p}_{\tilde{1}} + \hat{p}_{\tilde{2}} + \hat{p}_{\tilde{3}}) \rbrack ^2\rangle=2.91 \pm 0.06 $ \\
            $\langle \lbrack \Delta (\hat{x}_{\tilde{2}} - \hat{x}_{\tilde{3}}) \rbrack ^2\rangle+ \langle \lbrack \Delta (\hat{p}_{\tilde{1}} + \hat{p}_{\tilde{2}} + \hat{p}_{\tilde{3}}) \rbrack ^2\rangle=2.89 \pm 0.06$
        \end{tabular}
        \\\hline
        $3$-ii                                                        &
        GHZ state  $(\tilde{1}$, $\tilde{2}$, and $\tilde{3})$        & $0.896 \pm 0.007$ & $0.816 \pm 0.006$      &
        \begin{tabular}{c}
            $\langle \lbrack \Delta (\hat{x}_{\tilde{1}} - \hat{x}_{\tilde{2}}) \rbrack ^2\rangle+ \langle \lbrack \Delta (\hat{p}_{\tilde{1}} + \hat{p}_{\tilde{2}} + \hat{p}_{\tilde{3}}) \rbrack ^2\rangle=3.39 \pm 0.04$ \\
            $\langle \lbrack \Delta (\hat{x}_{\tilde{2}} - \hat{x}_{\tilde{3}}) \rbrack ^2\rangle+ \langle \lbrack \Delta (\hat{p}_{\tilde{1}} + \hat{p}_{\tilde{2}} + \hat{p}_{\tilde{3}}) \rbrack ^2\rangle=3.09 \pm 0.04$
        \end{tabular}
        \\\hline
        $3$-iii                                                       &
        GHZ state         $(\tilde{1}$, $\tilde{2}$, and $\tilde{3})$ & $0.888 \pm 0.008$ & $0.826 \pm 0.007$      &
        \begin{tabular}{c}
            $\langle \lbrack \Delta (\hat{x}_{\tilde{1}} - \hat{x}_{\tilde{2}}) \rbrack ^2\rangle+ \langle \lbrack \Delta (\hat{p}_{\tilde{1}} + \hat{p}_{\tilde{2}} + \hat{p}_{\tilde{3}}) \rbrack ^2\rangle=3.29 \pm 0.05$ \\
            $\langle \lbrack \Delta (\hat{x}_{\tilde{2}} - \hat{x}_{\tilde{3}}) \rbrack ^2\rangle+ \langle \lbrack \Delta (\hat{p}_{\tilde{1}} + \hat{p}_{\tilde{2}} + \hat{p}_{\tilde{3}}) \rbrack ^2\rangle=3.24 \pm 0.07$
        \end{tabular}
        \\\hline
        $4$-i                                                         &
        \begin{tabular}{c}
            Triangle cluster \\ state $(\tilde{1}$, $\tilde{2}$, and $\tilde{3})$
        \end{tabular}
                                                                      & $0.909 \pm 0.019$ & $0.863 \pm 0.015$      &
        \begin{tabular}{c}
            $\langle \lbrack \Delta (\hat{p}_{\tilde{1}} - \hat{x}_{\tilde{2}} - \hat{x}_{\tilde{3}}) \rbrack ^2\rangle+ \langle \lbrack \Delta (\hat{p}_{\tilde{3}} - \hat{x}_{\tilde{1}} - \hat{x}_{\tilde{2}}) \rbrack ^2\rangle=3.21 \pm 0.05$ \\
            $\langle \lbrack \Delta (\hat{p}_{\tilde{2}} - \hat{x}_{\tilde{1}} - \hat{x}_{\tilde{3}}) \rbrack ^2\rangle+ \langle \lbrack \Delta (\hat{p}_{\tilde{3}} - \hat{x}_{\tilde{1}} - \hat{x}_{\tilde{2}}) \rbrack ^2\rangle=3.80 \pm 0.04$
        \end{tabular}
        \\\hline
        $4$-ii                                                        &
        \begin{tabular}{c}
            Linear cluster \\ state $(\tilde{1}$, $\tilde{2}$, and $\tilde{3})$
        \end{tabular}                                    & $0.976 \pm 0.007$ & $0.920 \pm 0.008$      &
        \begin{tabular}{c}
            $\langle \lbrack \Delta (\hat{p}_{\tilde{1}}  - \hat{x}_{\tilde{3}}) \rbrack ^2\rangle+ \langle \lbrack \Delta (\hat{p}_{\tilde{3}} - \hat{x}_{\tilde{1}} - \hat{x}_{\tilde{2}}) \rbrack ^2\rangle=2.77 \pm 0.05$ \\
            $\langle \lbrack \Delta (\hat{p}_{\tilde{2}}- \hat{x}_{\tilde{3}}) \rbrack ^2\rangle+ \langle \lbrack \Delta (\hat{p}_{\tilde{3}} - \hat{x}_{\tilde{1}} - \hat{x}_{\tilde{2}}) \rbrack ^2\rangle=2.75 \pm 0.04$
        \end{tabular}
    \end{tabular}
    \caption{{Fidelities and inseparability parameters for the output modes of various three-mode linear operations. See text for details.}
    }
    \label{tab:fidelites}
\end{table*}
\normalsize

\textit{Experimental setup} ---We develop the dual-loop circuit with $N=3$ that can perform universal and {programmable} three-mode linear optical quantum operations, as shown in {Fig.~\ref{fig:loop}\blue{(b)}} ~\cite{supp}. Our setup achieves all the functionalities in the original proposal of the dual-loop circuit~\cite{motes2014scalable}. In our setup, we choose the time interval of $\tau=\SI{66}{ns}$ and the corresponding inner and outer loop lengths of \SI{19.8}{m} ($\tau$) and \SI{39.6}{m} ($(N-1)\tau$), respectively. Both the inner and outer loops are {phase-locked}. Two switches, one VPS, and one VBS are incorporated in the loops and synchronously controlled every $\tau=\SI{66}{ns}$. The adjustable range of VPS phase shift and VBS transmissivity covers the entire range required for universality from 0 to $2\pi$ and from 0 to 1, respectively. To evaluate the performance of the operations in the dual-loop circuit, three-mode squeezed-state pulses are injected and each output pulse is measured by a homodyne detector (HD) with a variable measurement basis $\hat{x}\cos \phi (t)+\hat{p}\sin \phi(t)$, where $\hat{x}$ and $\hat{p}$ are the quadrature operators of the light field and $\phi$ is called a homodyne angle. Our control sequence is based on Figs.~\ref{fig:3mode_ver2}\blue{(b)--(f)}, but the final unimportant local phase shifts $(\theta_4, \theta_5, \theta_6)$ in the VPS are omitted and equivalently performed by shifting the measurement bases at the HD. This reduces the number of round trips of optical pulses in the loops and thus minimizes the optical loss during the operations.

\textit{Experimental results} ---As a demonstration of programmable multi-mode linear optical operations in the time domain, we perform nine different three-mode operations on the input $p$-squeezed state pulses using our dual-loop circuit.
It is known that appropriate linear operations can transform such squeezed states into various multi-mode continuous-variable entangled states~\cite{takeda2019demand,vanLoock2007Building}.
Thus our overall system can also be regarded as a general multi-mode continuous-variable photonic entanglement synthesizer~\cite{takeda2019demand}.
We mainly adopt such operations for the demonstration and quantitatively evaluate the covariance matrices of the output states to verify the validity of the operations.
Note that the covariance matrices fully characterize the output states which are always Gaussian states with zero-mean quadratures in this experiment.
In addition, we evaluate the degree of entanglement of the generated entangled states to show that the operations are performed in the quantum regime.

As shown in Fig.~\ref{fig:3mode_ver2}, three-mode linear operations are composed of three two-mode BS interactions. First, we run our dual-loop circuit in the simplest setting where all these BS interactions are switched off by always setting the VBS transmissivity to 1 (Operation 1). The equivalent circuit in the path encoding is shown in the left panel of Fig.~\ref{fig:result_all}\blue{(a)}. This operation only rearranges the order of the input modes and thus each output mode becomes a $p$-squeezed state. The experimental output covariance matrix is shown in the middle panel of Fig.~\ref{fig:result_all}\blue{(a)}. As expected, it shows (anti-)squeezed variances in the $p$ ($x$) quadratures for all modes, while not showing correlation between these modes. The theoretical covariance matrix including estimated optical losses~\cite{supp} is also plotted in the right panel of Fig.~\ref{fig:result_all}\blue{(a)}, which reasonably well agrees with the experimental one.
As can be seen from the covariance matrix, the output modes are slightly asymmetric. This is because, in our sequence, the squeezed state coming to mode $\tilde{1}$ suffers from an extra round-trip loss in the outer loop compared to the other modes.

Next, we perform three-mode linear optical operations that generate various continuous-variable entangled states. {In particular}, we choose eight different operations and generate four types of entangled states: Einstein-Podolsky-Rosen (EPR) states~\cite{ou1992realization} generated by switching on one BS interaction (Operation 2-i, ii, iii), Greenberger-Horne-Zeilinger (GHZ) states~\cite{aoki2003experimental} generated by switching on two BS interactions (Operation 3-i, ii, iii), and two shapes of cluster states~\cite{su2007experimental} generated by switching on all three BS interactions (Operation 4-i, ii).
Figures~\ref{fig:result_all}\blue{(b)} and \ref{fig:result_all}\blue{(c)} are the representative results for Operations 3 and 4, showing the equivalent path-encoding circuits and the output covariance matrices.
As opposed to Operation 1, the experimental covariance matrices show non-zero off-diagonal elements for all cases, which implies that some of the modes are entangled.
In addition, the experimental covariance matrices agree well with the theoretical ones, demonstrating that the dual-loop circuit performs the three-mode operations as expected. The covariance matrices of all the other operations are summarized in Supplemental Material~\cite{supp}.

Finally, we quantitatively evaluate the performance of all the above nine operations.
We calculate and summarize the fidelities (Fidelity($\mathrm{I}$))  between the experimental output quantum states and the theoretical ones including losses in Table \ref{tab:fidelites}.
All the operations show reasonably high fidelities of $\sim 0.9$ or above.
The deviations between the experimental and theoretical results can be attributed to the unwanted phase drift or fluctuation in the loops as well as the deviation between the estimated losses and the actual ones.
Note that the fidelities (Fidelity($\mathrm{I}\hspace{-1.2pt}\mathrm{I}$)) between the experimental output states and ideal theoretical ones without including loop losses are also summarized in Table~\ref{tab:fidelites}.
We also assess inseparability parameters for the generated entangled states to quantify the degree of entanglement. For all the cases except for Operation 1,  the sufficient condition for full inseparability is that the inseparability parameter is below 4 ($\hbar = 2$)~\cite{van2003detecting}.
As summarized in Table \ref{tab:fidelites},
all the measured inseparability parameters satisfy this condition, indicating that all these operations are properly performed in the quantum regime.
Here the inseparability parameters are slightly worse than the corresponding values in our previous single-loop experiment~\cite{takeda2019demand} due to the additional loss introduced by the extra round trip in the outer loop.
Note that all these operations in our dual-loop circuit are performed without any changes to the hardware configuration.
Thus, these results demonstrate the validity, programmability, and deterministic operation of our dual-loop circuit that is universal for three-mode linear optical operations.

    {
        \textit{Discussion} ---In conclusion, we developed a scalable dual-loop circuit with complete dynamic controllability to perform universal three-mode linear optical operations in the time domain. We showed its applicability to universal QIP in the continuous-variable regime. Furthermore, since our circuit can deal with any input state including qubits, it is also applicable to the qubit regime.
        The number of processable modes can be scaled up by several orders of magnitude either by using a km-long optical fiber for a stable and longer outer loop with comparable losses or by using broader-bandwidth light sources and electronics to shorten the time interval of pulses~\cite{Takeda2019Toward}. Furthermore, our dual-loop circuit can be integrated with other quantum light sources pumped by either continuous-wave or pulsed lasers. This work is extendable to loop-based universal optical quantum computers by incorporating feedforward systems~\cite{Rohde2015Simple,takeda2017universal}, and thus a crucial step toward large-scale universal optical QIP.

    }
    
    {
        \textit{Note added.} ---We have recently become aware of a work~\cite{Shang2022Universal} in which a loop circuit with a different configuration performed universal linear optical operations in the time domain. However, in the same way as the previous works~\cite{he2017time, schreiber20122d}, this work was designed for a specific non-universal task (Gaussian boson sampling) and sampled the output at a fixed measurement basis without characterizing the operations themselves.
    }

\begin{acknowledgments}
    This work was partly supported by JSPS KAKENHI Grant Numbers 20H01833 and 21K18593, MEXT Leading Initiative for Excellent Young Researchers, Toray Science Foundation (19-6006), and the Canon Foundation. The authors thank Akira Furusawa for providing space for the experiment. The authors also thank Takahiro Mitani for the careful {proofreading} of the manuscript.
\end{acknowledgments}

\bibliography{reference}
\nocite{herriott1964off,Black2001introduction, simon1994quantum, folland2016harmonic, de2018simple, clements2016optimal, fukui2018high, yoshikawa2016invited, motes2015implementing, wang2018toward}
\end{document}


\title{Supplemental Material for \\
    Time-domain universal linear optical operations \\
    for universal quantum information processing
}

\author{Kazuma Yonezu}

\author{Yutaro Enomoto}%

\author{Takato Yoshida}%

\author{Shuntaro Takeda}%
\email{takeda@ap.t.u-tokyo.ac.jp}

\date{\today}

\maketitle

\section*{Experimental setup}

\noindent
{Figure~\blue{1(b)}} illustrates our experimental setup for the dual-loop circuit with $N=3$. This setup is extended from our previous single-loop circuit, which is described in detail in Refs.~\cite{takeda2019demand, enomoto2021programmable}. A squeezed-vacuum beam at 860 nm is generated from a {continuously pumped} optical parametric oscillator (OPO) with a bandwidth of $\sim\SI{80}{\MHz}$, and the input squeezed-state pulses are defined at the time interval of $\tau = \SI{66}{ns}$. These pulses are then sent to the dual-loop circuit. The inner and outer loops are constructed by Herriott-type optical delay lines~\cite{herriott1964off} and have round-trip times of $\tau = \SI{66}{ns}$ (\SI{19.8}{m}) and $(N-1) \tau = \SI{132}{ns} $ (\SI{39.6}{m}), respectively. This dual-loop circuit includes four dynamically controllable elements: Switch-1, Switch-2, VPS, and VBS.
For example, the VBS is composed of an electro-optic modulator (EOM) named EOM-2 and two polarizing BSs (PBSs). Additionally, a quarter-wave plate (QWP) is inserted between the PBSs to set the VBS transmissivity to 0.5 by default for phase locking of the loops. We then apply appropriate voltages to EOM-2 to dynamically control the transmissivity. The maximum output voltage of the driver for EOM-2 is 2.2 kV, which is sufficient to set an arbitrary beam splitter transmissivity from 0 and 1 under the measured half-wave voltage of 0.88 kV. Switch-1 (Switch-2) works in the same way by using EOM-1 (EOM-3). The VPS is also realized by EOM-4. The maximum output voltage of the driver for EOM-4 is 2.2 kV, which is sufficient to set an arbitrary amount of phase shift from 0 to 2$\pi$ under the measured coefficient of 0.26 kV rad$^{-1}$. Our EOM drivers can apply freely chosen multiple voltages on the VBS and VPS in a similar way as in Ref.~\cite{enomoto2021programmable}, which enables the implementation of an arbitrary linear optical operation. The output pulses from the dual-loop circuit are finally measured by the HD, whose homodyne angle is controlled by EOM-5. All the above EOMs are synchronously and dynamically controlled every $\tau = \SI{66}{ns}$ by the timing controller. The switching time of these EOMs is $\sim$ \SI{10}{ns}.

Switch-1 and Switch-2 introduce an extra \ang{180} phase shift depending on their working conditions~\cite{takeda2019demand}. In our setting, when Switch-1 transmits an incoming mode from the outer loop, the mode suffers from a \ang{180} phase shift. Therefore, in the path-encoding circuits of {Figs.~\blue{2(a)} and \blue{3}}, the \ang{180} phase shift is added to the path corresponding to the outer loop. On the other hand, when Switch-2 exports a mode in the circuit into the HD, it gives a \ang{180} phase shift to the mode. Since all of the output modes, which are measured by the HD, commonly suffer from this phase shift, it does not affect the relative phase shift between these modes, and thus this effect can be ignored.

\section*{Data analysis}

\noindent
The quadratures of the output modes are extracted by applying temporal mode functions to the acquired homodyne signal.
The function for the $k$-th mode ($k=1,2,3$) is defined as
\begin{equation}
    f_k(t) \propto
    \begin{cases}
        e^{-\gamma^2 (t-t_k)^2} (t-t_k) & (2|t-t_k| \leq \Delta t) \\
        0                               & (\mathrm{otherwise})
    \end{cases}
\end{equation}
and normalized to be $\int_{-\infty}^{\infty}|f_k(t)|^2 dt = 1$~\cite{takeda2019demand, enomoto2021programmable}.
Here we choose $\Delta t=\SI{46}{ns}$, $\gamma = \SI{6 E7}{/s}$, and $t_k = t_0 + (k-1)\tau$. This mode function has no low-frequency components and thus can reduce the unfavorable effect of a high-pass filter in our measurement system~\cite{yoshikawa2016invited}. The parameters of the mode function are chosen so that the frequency spectrum of the mode function is within the bandwidth of the OPO.
To calculate the output covariance matrices and inseparability parameters for each operation, we acquire 5000 samples of the output quadratures for each of the following five different measurement bases:
$(\hat{p}_{\tilde{1}}, \hat{x}_{\tilde{2}}, \hat{x}_{\tilde{3}})$, $(\hat{x}_{\tilde{1}}, \hat{p}_{\tilde{2}}, \hat{x}_{\tilde{3}})$, $(\hat{x}_{\tilde{1}}, \hat{x}_{\tilde{2}}, \hat{p}_{\tilde{3}})$, $(\hat{p}_{\tilde{1}}, \hat{p}_{\tilde{2}}, \hat{p}_{\tilde{3}})$, and
$(\frac{\hat{x}_{\tilde{1}}+\hat{p}_{\tilde{1}}}{\sqrt{2}}, \frac{\hat{x}_{\tilde{2}}+\hat{p}_{\tilde{2}}}{\sqrt{2}}, \frac{\hat{x}_{\tilde{3}}+\hat{p}_{\tilde{3}}}{\sqrt{2}})$.
Here, $\hat{x}$, $\frac{\hat{x}+\hat{p}}{\sqrt{2}}$, and $\hat{p}$ for each mode can be measured by adding phase shifts of \ang{0}, \ang{45}, and \ang{90} to the local oscillator (LO) beam, respectively.
All the elements of the output covariance matrices as well as the values of the inseparability parameters can be calculated from the above five sets of measurement results.

\section*{Dual-loop phase locking}

\noindent
Let us here describe the way to lock the phases of the dual loops that can be regarded as coupled resonators.
For the phase locking, a reference light is injected into the OPO after being phase-modulated at two frequencies of \SI{300}{kHz} and \SI{4.5}{MHz}. This light is then detected at an unused port of the second PBS in Fig.~{\blue{1(b)}}. We demodulate the detection signal at \SI{300}{kHz} and \SI{4.5}{MHz}, thereby obtaining the error signals for the outer and inner loops, respectively. The outer and inner loops can be simultaneously phase-locked by feeding back these error signals.
Here, since the Pound--Drever--Hall error signal of a resonator behaves differently for lower and higher frequencies than the linewidth of the resonator~\cite{Black2001introduction}, the demodulated signals contain the error signals of the inner and outer loops in different proportions.
These specific frequencies are chosen so that the error signals of the two loops are separated the most based on numerical simulations.


\begin{figure}[tb]
    \centering
    \includegraphics[scale = 0.43]{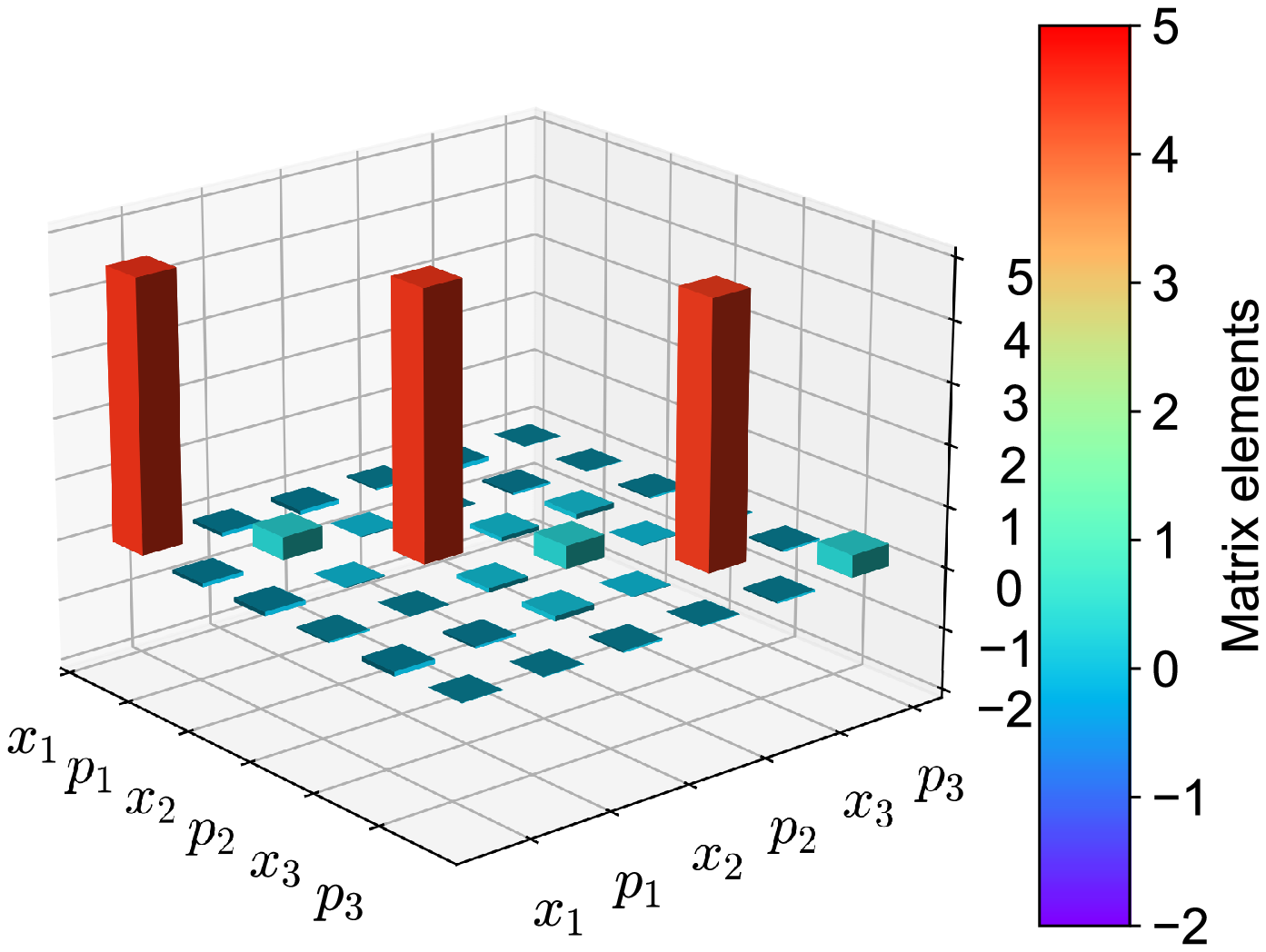}
    \caption{Measured covariance matrix of the three input modes of squeezed vacuum states.}
    \label{fig: Sqwoeom}
\end{figure}

\section*{Optical loss}

\noindent
Optical losses in our setup are estimated from two measurement results.
First, optical losses in the generation and measurement part of the squeezed light are estimated by the input covariance matrix in Fig.~\ref{fig: Sqwoeom}. This matrix is measured by keeping the VBS transmissivity 0 and thereby directly transmitting the input squeezed modes to the output port without letting them go around the loops. In this case, all the squeezed modes suffer from the same amount of optical losses, including the OPO internal loss, the propagation loss from the OPO to the HD, and the readout loss of the HD.
From this input covariance matrix, the total loss is estimated to be \SI{23}{\%}, and the initial pure squeezing level before suffering from any optical losses is estimated to be \SI{7.4}{dB}.
Next, the round-trip optical losses in the loops are estimated from the output covariance matrix of Operation 1 in {Fig.~\blue{3(a)}}.
Here the squeezed states coming to mode $\tilde{2}$ and $\tilde{3}$ suffer from the round-trip loss in the inner loop in addition to the losses in the generation and measurement part. Furthermore, the squeezed state coming to mode $\tilde{1}$ suffers from the extra round-trip loss in the outer loop. From these facts, the round-trip losses in the inner and outer loops are estimated to be \SI{15}{\%} and \SI{20}{\%}, respectively.
The theoretical output covariance matrices are calculated by using the estimeted squeezing level and including the effect of the estimated losses.
\noindent
In our dual-loop circuit, the number of modes can be increased by using longer loops, but it increases the round-trip losses in the loops. One of the optimal options to realize stable and longer loops is to use optical fibers, which introduce \SI{0.2}{\dB / \km} loss at the minimum. On the other hand, scaling up universal QIP requires the round-trip loss to be below a fault-tolerant threshold for quantum error correction. For example, a theoretical work reported that fault-tolerant quantum computing is possible with Gottesman-Kitaev-Preskill (GKP) qubits of $\sim\SI{10}{\dB}$ squeezing ~\cite{fukui2018high}. \SI{10}{\dB} squeezing is possible with up to \SI{10}{\%} losses, thus \SI{10}{\%} can be considered a loss threshold for error correction in this case. This loss threshold limits the fiber length below a few kilometers. However, the number of modes can also be increased by several orders by instead shortening the pulse interval~\cite{Takeda2019Toward}. More detailed analyses of how losses scale with the number of modes have been addressed in Refs.~\cite{motes2015implementing, wang2018toward}.

\section*{Multi-mode linear optical operation}

\noindent
Here we define general multi-mode linear operations
and explain how covariance matrices are transformed by such operations.
In the Heisenberg picture, the input-output relation of $N$-mode linear operations {is} defined in a matrix form as
\begin{equation}
    \begin{pmatrix}
        \hat{a}_1^{\mathrm{out}} \\
        \hat{a}_2^{\mathrm{out}} \\
        \vdots                   \\
        \hat{a}_N^{\mathrm{out}}
    \end{pmatrix}
    = U_N
    \begin{pmatrix}
        \hat{a}_1^{\mathrm{in}} \\
        \hat{a}_2^{\mathrm{in}} \\
        \vdots                  \\
        \hat{a}_N^{\mathrm{in}}
    \end{pmatrix},
\end{equation}
where $\hat{a}_{i}^{k}$ ($k = \text{in}, \text{out}$) is an input or output annihilation operator of mode $i$
and $U_N$ is a unitary matrix.
This relation can be reformed by using quadratures
$\hat{x}_{i}^k=\hat{a}_{i}^k + (\hat{a}_{i}^k)^\dagger$ and $\hat{p}_{i}^k = -i\hat{a}_{i}^k + i(\hat{a}_{i}^k)^\dagger$
as
\begin{equation}
    \hat{\xi}^{\mathrm{out}} = (WA)^{-1}
    \begin{pmatrix}
        U_N & 0     \\
        0   & U_N^*
    \end{pmatrix}
    WA \hat{\xi}^{\text{in}}
    \equiv
    S\hat{\xi}^{\text{in}},
\end{equation}
where $\hat{\xi}^k = (\hat{x}_1^k, \hat{p}_1^k, \cdots \hat{x}_N^k, \hat{p}_N^k)^T$,
$A$ and $W$ are $2N$-dimensional matrices defined by
\begin{equation}
    \begin{split}
        A_{ij}&=
        \begin{cases}
            1 & (j \text{ even and } i = N+j/2)  \\
            1 & (j \text{ odd and } i = (j+1)/2) \\
            0 & (\text{otherwise})
        \end{cases},\\
        W&=
        \begin{pmatrix}
            I & iI  \\
            I & -iI
        \end{pmatrix},
    \end{split}
\end{equation}
and $I$ is an $N$-dimensional identity matrix~\cite{simon1994quantum,folland2016harmonic}.
This quadrature transformation also changes the corresponding covariance matrix $\Gamma^k$,
whose elements are defined as
$\Gamma_{ij}^k = \langle \hat{\xi}_i^k \hat{\xi}_j^k + \hat{\xi}_j^k \hat{\xi}_i^k \rangle /2 - \langle \hat{\xi}_i^k \rangle \langle \hat{\xi}_j^k \rangle$.
The input-output relation of the covariance matrices can be written as
\begin{equation}
    \Gamma^{\mathrm{out}} = S \Gamma^{\mathrm{in}} S^T.
\end{equation}
This transformation rule is used to derive the theoretical covariance matrices.

\section*{$N$-mode linear optical operation in dual loops}

\begin{figure}[tbp]
    \centering
    \includegraphics[scale = 1]{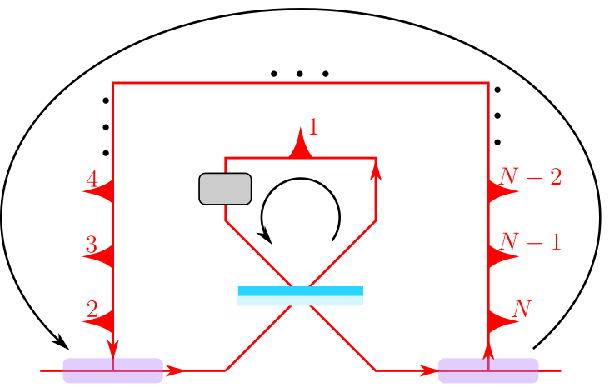}
    \caption{{Universal $N$-mode linear optical operations in the dual-loop circuit.}
        Mode $1$ is in the inner loop and the other modes are in the outer loop.
        Two-mode BS operations are repeatedly performed between the modes in the inner and outer loops.
    }
    \label{fig:loop_theory}
\end{figure}

\noindent
We show the general sequence to perform an arbitrary $N$-mode linear optical operation in the dual-loop circuit.
In general, an arbitrary linear optical operation $U_N$ can be decomposed into sequences of BS and PS operations. Several ways for the decomposition are known~\cite{reck1994experimental, de2018simple, clements2016optimal}, but here we drive a slightly modified decomposition that is compatible with the dual-loop circuit.
When the $N$-dimensional unitary matrix $U_N$ is multiplied with appropriate unitary matrices $T_{l,m}^{(1)}$, the effective dimension of the unitary matrix can be reduced as
\begin{equation}
    U_N T_{1,2}^{(1)} T_{1,3}^{(1)} \cdots T_{1, N}^{(1)} =
    \begin{pmatrix}
        U_{N-1} & 0             \\
        0       & e^{i\alpha_N}
    \end{pmatrix},
\end{equation}
where $\alpha_N$ is a real constant,
$U_{N-1}$ is a certain $(N-1)$-dimensional unitary matrix,
and $T_{l,m}^{(1)}$ is an $N$-dimensional identity matrix with the $(l,l)$, $(l, m)$, $(m,l)$, and $(m,m)$ elements replaced by $e^{i\phi_{l,m}}\sin \omega_{l,m} $, $-e^{i\phi_{l,m} }\cos{\omega_{l,m} }$, $\cos \omega_{l,m}$, and $\sin \omega_{l,m}$, respectively~\cite{reck1994experimental}.
The matrix $U_N$ can be {diagonalized} by repeating this procedure as
\begin{equation}
    U_N T_{1,2}^{(1)} T_{1,3}^{(1)} \cdots T_{1, N}^{(1)}
    T_{1,2}^{(2)}T_{1,3}^{(2)} \cdots T_{1, N-1}^{(2)}
    \cdots
    T_{1,2}^{(N-1)}
    =D,
\end{equation}
where $D=\text{diag}(e^{i\alpha_1}, e^{i\alpha_2}, \cdots, e^{i\alpha_N})$ for real constants $\alpha_1, \alpha_2, \cdots, \alpha_N$.
As a result, $U_N$ can be decomposed as
\begin{equation}
    \begin{split}
        U_N &= D (T_{1,2}^{(1)} T_{1,3}^{(1)} \cdots T_{1, N}^{(1)}
        T_{1,2}^{(2)}T_{1,3}^{(2)} \cdots T_{1, N-1}^{(2)}
        \cdots
        T_{1,2}^{(N-1)})^{-1} \\
        &= D( T_{1, 2}^{(N-1)})^{-1} \cdots( T_{1, N-1}^{(2)})^{-1} \\
        & \qquad \cdots (T_{1, 3}^{(2)})^{-1} (T_{1, 2}^{(2)})^{-1} (T_{1, N}^{(1)})^{-1} \cdots (T_{1,3}^{(1)})^{-1} (T_{1, 2}^{(1)})^{-1}.
    \end{split}
    \label{eq:unitary}
\end{equation}
This decomposition has clear correspondence with the dual-loop circuit and thus is suitable for the implementation in the circuit. In the dual-loop circuit, mode 1 is stored in the inner loop, while the other modes are stored in the outer loop, as shown in Fig.~\ref{fig:loop_theory}. Then mode 1 is sequentially phase-shifted and interfered with mode $2,3 \cdots, N$ in the outer loop. Such sequential operations can perform
$(T_{1, N}^{(1)})^{-1} \cdots (T_{1,3}^{(1)})^{-1} (T_{1, 2}^{(1)})^{-1}$.
During the next round trip, we can perform
$( T_{1, N-1}^{(2)})^{-1} \cdots (T_{1, 3}^{(2)})^{-1} (T_{1, 2}^{(2)})^{-1}$
(the unnecessary term $( T_{1, N}^{(2)})^{-1}$ can be skipped by setting the VBS transmissivity to 0). By repeating this sequence, the dual-loop circuit can perform
$(T_{1, 2}^{(N-1)})^{-1} \cdots( T_{1, N-1}^{(2)})^{-1} \cdots (T_{1, 3}^{(2)})^{-1} (T_{1, 2}^{(2)})^{-1} (T_{1, N}^{(1)})^{-1} \cdots (T_{1,3}^{(1)})^{-1} (T_{1, 2}^{(1)})^{-1}$.
Finally, individual phase shifts are applied to all modes by keeping the VBS transmissivity $1$, which implements the diagonal matrix $D$ and completes the $N$-mode linear optical operation in Eq.~(\ref{eq:unitary}). We perform three-mode linear optical operations based on the above decomposition.

In general, an $N$-mode linear optical operation with our dual-loop circuit requires individual $N-1$ modes except for mode 1 to circulate the outer loop for $N-1$ times, as indicated in Eq.~(\ref{eq:unitary}). Thus, larger-scale circuits inevitably suffer from more loss due to the extra $N-1$ round trip losses in the outer loop.

\newpage
\section*{Supplementary results}
\begin{figure}[htbp]
    \centering
    \includegraphics[scale=1]{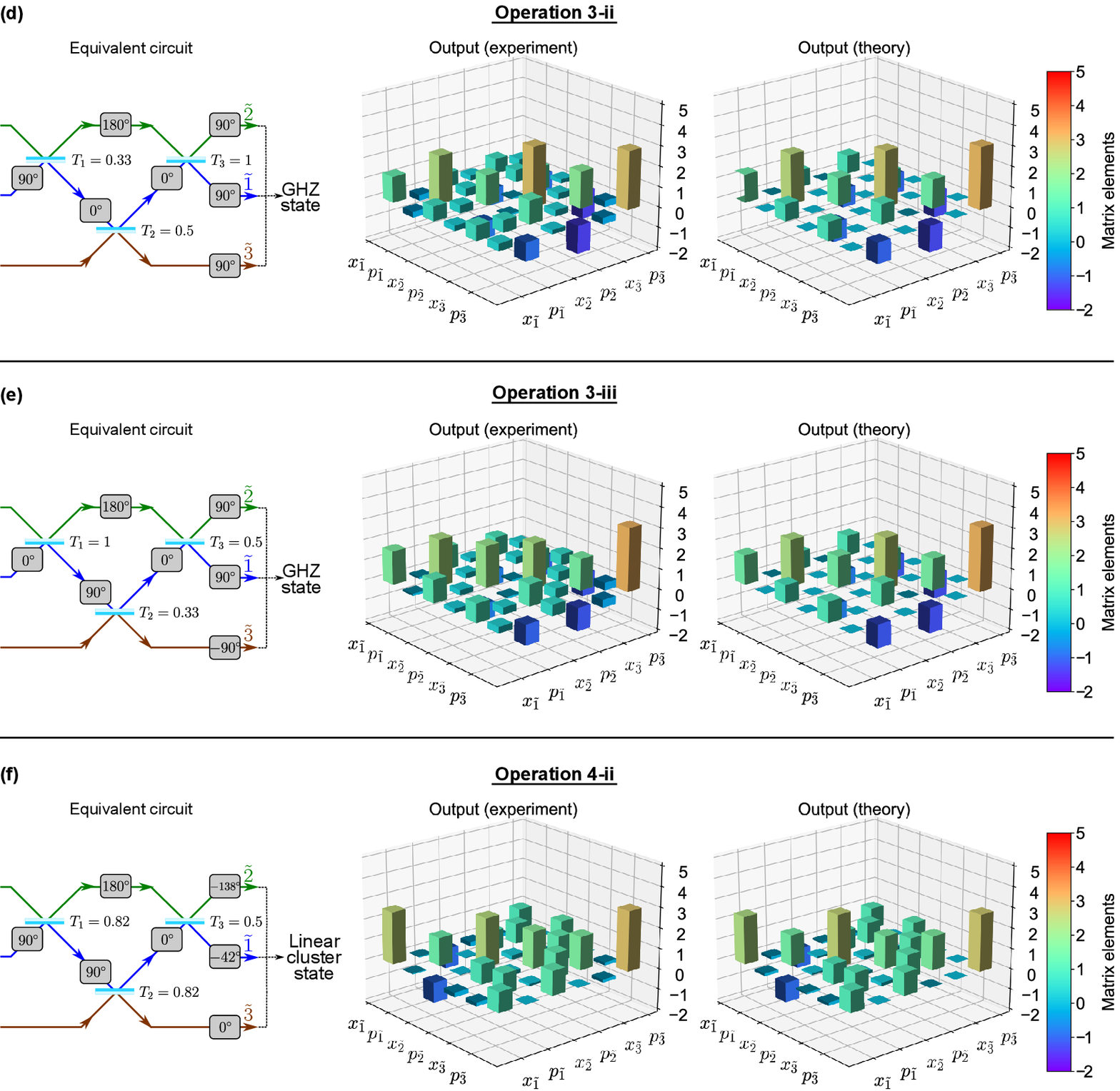}
    \caption{
    {Experimental results of three-mode linear operations in the dual-loop circuit.}
    The left column shows the equivalent circuits in the path encoding.
    The middle and right columns show the experimental and theoretical output covariance matrices, respectively.
    The matrix elements represent covariances
    $\langle \hat{\xi}_i \hat{\xi}_j + \hat{\xi}_j \hat{\xi}_i \rangle/2 - \langle \hat{\xi}_i \rangle \langle \hat{\xi}_j\rangle$,
    where $\braket{\cdots}$ denotes the mean value and
    $\hat{\xi}=(\hat{x}_{\tilde{1}}, \hat{p}_{\tilde{1}}, \hat{x}_{\tilde{2}}, \hat{p}_{\tilde{2}}, \hat{x}_{\tilde{3}}, \hat{p}_{\tilde{3}})^T$.
    The vacuum variance is set to 1 ($\hbar = 2$).
    }
    \label{fig:supplementary}
\end{figure}

The experimental results of the three-mode linear optical operations not shown in the main text are summarized in Fig.~\ref{fig:supplementary}.
\begin{figure}[htbp]
    \centering
    \includegraphics[scale=1]{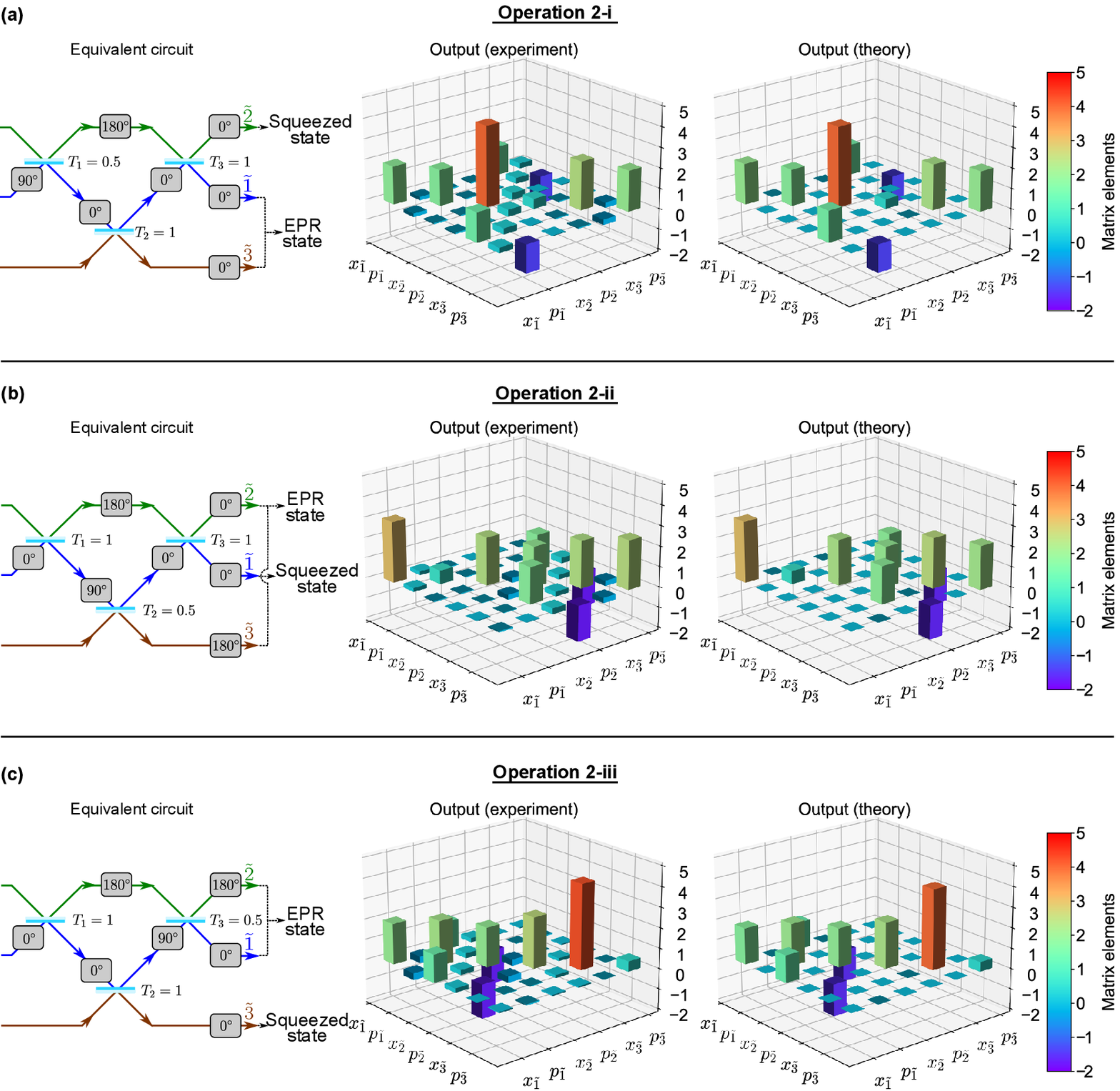}
\end{figure}

\newpage
\bibliography{reference}